\documentclass[aps,pra,notitlepage,amsmath,amssymb,amsfonts,superscriptaddress]{revtex4-1}

\usepackage{graphicx}
\usepackage{subfigure}
\usepackage{amsthm}
\usepackage{verbatim}
\usepackage{dcolumn}
\usepackage{bm}
\usepackage{epsf}
\usepackage{color}
\usepackage[outdir=./]{epstopdf}
\usepackage[colorlinks=true,citecolor=blue,linkcolor=blue,urlcolor=blue]{hyperref}

\newcommand{\bra}[1]{\left\langle #1\right|}
\newcommand{\ket}[1]{\left|#1\right\rangle}
\newcommand{\braket}[2]{\left\langle #1|#2\right\rangle}

\newcommand{\tr}[1]{\mathrm{tr}\left\{#1\right\}}

\newcommand{\la}{\left\langle}
\newcommand{\ra}{\right\rangle}

\newcommand{\td}{\mathrm{d}}

\newcommand{\e}[1]{\exp{\left(#1\right)}}

\newcommand{\id}{\mathbb{I}}

\newcommand{\bla}{bla\\bla\\bla\\bla\\bla}

\begin{document}
	
\title{Non-hermitian quantum thermodynamics}
\author{Bart\l{}omiej Gardas}
\email{bartek.gardas@gmail.com}
\affiliation{Theoretical Division, Los Alamos National Laboratory, Los Alamos, NM 87545, USA}
\affiliation{Institute of Physics, University of Silesia, 40-007 Katowice, Poland}
\author{Sebastian Deffner}
\author{Avadh Saxena}
\affiliation{Theoretical Division, Los Alamos National Laboratory, Los Alamos, NM 87545, USA}
\affiliation{Center for Nonlinear Studies, Los Alamos National Laboratory, Los Alamos, NM 87545, USA}

\begin{abstract}
	Thermodynamics is the phenomenological theory of  heat and work. Here we analyze to what extent quantum thermodynamic
	relations are immune to the underlying mathematical formulation of quantum mechanics. As a main result, we show that 
	the Jarzynski equality holds true for all non-hermitian quantum systems with real spectrum. This equality expresses
	the second law of thermodynamics for isothermal processes arbitrarily far from equilibrium. In the quasistatic limit 
	however, the second law leads to the Carnot bound which is fulfilled even if some eigenenergies are complex provided
	they appear in conjugate pairs. Furthermore, we propose two setups to test our predictions. Namely with strongly 
	interacting excitons and photons in a semiconductor microcavity and in the non-hermitian tight-binding model.
\end{abstract}

\date{\today}

\maketitle
	
\vspace{0.5cm}
{\noindent  \bf \Large Introduction}
\vspace{0.5cm}

\noindent 
More and more non-hermitian systems are becoming experimentally accessible~\cite{relav}. Therefore, it has become evident that questions
concerning foundations of quantum mechanics are no longer only of academic interest. Recent experiments have demonstrated that hermiticity
may not be as fundamental as mandated by quantum mechanics~\cite{Dirac1939,VonNeumann1955}.  For instance, in~\cite{optic} a spontaneous 
$\mathcal{P}\mathcal{T}$-symmetry breaking has been observed indicating a condition weaker than hermiticity (namely $\mathcal{P}\mathcal{T}$~\cite{Bender1998}) being realized in nature. Furthermore, in~\cite{gao} exceptional eigenenergies of complex value
have been measured challenging the reality of the spectrum imposed by hermiticity.

Conventional quantum mechanics is built upon the Dirac-von Neumann axioms~\cite{Dirac1939,VonNeumann1955}. These state that if $\mathcal{H}$
is a complex Hilbert space of countable, infinite dimension, then (i) observables of a quantum system are defined as hermitian operators $O$
on $\mathcal{H}$, (ii) quantum states $\ket{\phi}$ are unit vectors in $\mathcal{H}$, and (iii) the expectation value of an observable $O$ in
a state $\ket{\phi}$ is given by the inner product, $\la O\ra =\braket{\phi}{O\,\phi}$. Interestingly, only axioms (ii) and (iii) are of 
mathematical necessity needed for a proper probabilistic, physical theory. To demand, however, that any quantum mechanical theory has to be 
built on hermitian operators is rather mathematically convenient than being fundamentally necessary~\cite{Bender1998,Meng15}.

In particular, the restriction to hermitian observables excludes the description of, for instance, quantum field theories with $\mathcal{PT}$-symmetry, cases where the language of quantum mechanics is used for problems within classical statistical mechanics or diffusion in biological systems, or cases where effective complex potentials are introduced to describe interactions at edges \cite{Moiseyev2011}. Particularly striking examples are optical systems with complex index of refraction. Imagine, for instance, polarized light in a stratified, nontransparent, biaxially anisotropic, dielectric medium warped cyclically along the propagation direction. For such systems it has been shown~\cite{Berry2011} that not only a non-hermitian description becomes necessary, but also that physical intuition has to be invoked carefully. For instance, Berry highlighted \cite{Berry2011} that adiabatic intuition can be countered dramatically for systems with non-hermitian Hamiltonians.

Very recently, it has become evident that for a special class of non-hermitian systems, namely in $\mathcal{PT}$-symmetric quantum mechanics~\cite{brody}, the quantum Jarzynski equality holds without modification~\cite{Deffner2015a}. For isolated quantum systems
evolving under unitary dynamics the so-called two-time energy measurement approach has proven to be practical and powerful. In this 
paradigm, quantum work is determined by projective energy measurements at the beginning and the end of a process induced by an externally
controlled Hamiltonian. 
The Jarzynski equality~\cite{Jarzynski1997} together with subsequent Nonequilibrium Work Theorems, such as the Crooks fluctuation theorem \cite{Crooks1999}, is undoubtedly among the most important breakthroughs in modern Statistical Physics~\cite{OrtizdeZarate2011}. Jarzynski 
showed that for isothermal processes the second law of thermodynamics can be formulated as an \emph{equality}, no matter how far from equilibrium the system is driven~\cite{Jarzynski1997}, $\la \e{-\beta W}\ra=\e{-\beta \Delta F}$. Here $\beta$ is the inverse temperature of the environment, and $\Delta F$ is the free energy difference, \emph{i.e.}, the work performed during an infinitely slow process.  The angular brackets denote the average over an ensemble of  finite-time realizations of the process characterized 
by their nonequilibrium work $W$. 

The present study is dedicated to an even more fundamental question. In the following we will analyze to what extent quantum thermodynamic relations are immune to the underlying mathematical formulation of quantum mechanics. Contrary to different studies (see \emph{e.g.}~\cite{Deffner2015a}) conducted on a similar subject we present the \emph{broadest} possible class of non-hermitian systems that still allows a thermodynamic theory 
in the ``conventional" sense.

As a main result we will prove that equilibrium as well as non-equilibrium identities of quantum thermodynamics hold, without modification also for quantum systems described by pseudo-hermitian Hamiltonians~\cite{ali_diag}. Those systems have either entirely real spectrum or complex eigenvalues appear in complex conjugate pairs. In particular, we will show that  the Carnot statement of the second law of thermodynamics holds for any such system and that the quantum Jarzynski equality is not violated as long as the eigenvalue spectrum is real. If the two-time energy measurement could
be realized \emph{e.g.} in a microcavity~\cite{gao}, then the Jarzynski equality for pseudo-hermitian systems could be put into a test (see \emph{Discussion}). 

\vspace{0.5cm}
{\noindent \bf \Large Fundamentals of pseudo-hermitian quantum mechanics}
\vspace{0.5cm}

To address physical properties of recent experiments~\cite{gao,optic} we start by briefly reviewing the mathematical foundations of 
pseudo-hermitian quantum mechanics~\cite{Moiseyev2011}. Let $H$ be a general, non-hermitian Hamiltonian of a physical system, and we
assume for the sake of simplicity that the spectrum of $H$,  $\{E_n\}$, is discrete (possibly degenerate). Such a Hamiltonian is of 
physical relevance only if it is measurable, \emph{i.e.}, if a representation of the eigenbasis $\ket{E_{n,\alpha}}$ is experimentally
accessible. Then $H$ is diagonal in this basis. Here $n$ is the quantum number and $\alpha$ counts possible degeneracy. Diagonalizability
of $H$ is \emph{equivalent} to the existence of biorthonormal set of left, $\ket{\phi_{n,\alpha}} $, and right, $\ket{\psi_{n,\alpha}}$, 
eigenvectors~\cite{brody2}. In general, the energy eigenvalues are complex, and the eigenvalue problem reads~\cite{ali_diag}

\begin{equation}
\label{eq01}
H\ket{\psi_{n,\alpha}} = E_n\ket{\psi_{n,\alpha}}, \quad H^{\dagger}\ket{\phi_{n,\alpha}} = E_n^*\ket{\phi_{n,\alpha}},
\end{equation}
with $\braket{ \psi_{n,\alpha} }{ \phi_{m,\beta}}=\delta_{mn}\delta_{\alpha\beta}$ and $\sum_{n,\alpha}\ket{\psi_{n,\alpha}}\bra{\phi_{n,\alpha}}=\id$. 
A non-hermitian Hamiltonian such as~(\ref{eq01}) is called pseudo-hermitian if a $g$ exists such that
\begin{equation}
\label{eq04}
H^{\dagger} = g H g^{-1}
\quad
\text{and}
\quad 
g=g^{\dagger}.
\end{equation}
It does exist \emph{if and only if} either all eigenenergies are real \emph{or} complex ones appear in conjugate pairs with the same
degeneracy~\cite{ali_diag}. If none of those criteria are met $H$ is generally non-hermitian~\cite{Moiseyev2011}; yet it still can be
useful, \emph{e.g.} for an effective description of open quantum systems~\cite{Huelga}. However, when heat is exchanged the two-time 
energy measurement can no longer describe the work done during a thermodynamic process. Therefore we shall not focus on such cases 
here. Another interesting class relates to systems that interact with environments, but do not exchange heat. This phenomenon is called dephasing
(loss of information)~\cite{Alicki04}. For such systems, work can still be determined by the two-time energy measurement and the Jarzynski
equality holds as well~\cite{Kim15,Deffner12,Rastegin13}. 

Condition~\eqref{eq04} assures that $H$ is, in  fact, hermitian however with respect to a new inner product, namely 
\begin{equation}
\label{inner}
\braket{\psi}{\phi}_g := \langle\psi, g \phi\rangle.
\end{equation}
Note that $g$ always exists such that $ \braket{\psi}{\phi}_g$ is positive-definite (this is a genuine inner product), and it can be found 
\emph{if and only if} the spectrum of $H$ is real. To make a consistent definition of work for a quantum system within the two--time energy
measurement paradigm its spectrum has to be real. Therefore, unless stated otherwise, we shall always assume this to be the case. Then, 
Eq.~(\ref{eq04}) can be fulfilled by the following positive-definite operators ($g$ is a proper metric operator)~\cite{ali_unique}

\begin{equation}
\label{eq05}
g = \sum_{n,\alpha} \ket{\phi_{n,\alpha}}\bra{\phi_{n,\alpha}},
\quad
g^{-1} = \sum_{n,\alpha} \ket{\psi_{n,\alpha}}\bra{\psi_{n,\alpha}}.
\end{equation}
Often, $g$ fulfilling~(\ref{eq04}) can be deduced easily from physical properties such as the parity reflection or time reversal~\cite{cho}. 
Nevertheless, only Eq.~(\ref{eq05}) assures that $\braket{\psi}{\psi}_g>0$ for all states $\psi\not=0$. This means that the proper metric may
reflect ``symmetries" that are hidden from the observer~\cite{hidden,yeo}. For instance, if a rotation $V$ exists such that $V^{-1}HV$ is 
diagonal in an orthonormal basis, then $g=V^{\dagger}V$. This follows directly from Eq.~(\ref{eq05}). The last formula is especially useful 
in practice. It allows one to find the metric by analyzing an experimental setup (\emph{e.g.} inspecting the orientation of the axis, etc.). 

In the following we only consider cases where changes of the Hamiltonian are induced by a time--dependent thermodynamic process $\lambda_t$,
that is to say $H_t=H(\lambda_t)$. If such changes occur then the metric operator satisfying Eq.~(\ref{eq04}) is time-dependent. Nevertheless,
the dynamics is still governed by a time-depended Schr\"odinger equation. However, a slight modification becomes necessary to preserve 
unitarity~\cite{Znojil,gong},

\begin{equation}
\label{se}
i\hbar\partial_t U_t = (H_t + G_t)U_t , \quad G_t=-\frac{i\hbar}{2} g_t^{-1} \partial_tg_t.
\end{equation}
Above, $\partial_t $ denotes the derivative with respect to time $t$.  The Schr\"odinger equation~(\ref{se}) can also be rewritten in the standard 
form, that is, with $H_t$ being the generator. Indeed, it is sufficient to replace $\partial_t$ with a covariant derivative $D_t:=\partial_t+g_t^{-1} \partial_tg_t/2$~\cite{cov}. By construction the unique solution to Eq.~(\ref{se}) obeys the relation
\begin{equation}
\label{gU}
U_t^{\dagger}g_t=g_oU_t^{-1}, 
\quad
\text{where}
\quad
g_0:=g_{t=0}.
\end{equation}
This relation can be viewed as the corresponding unitarity condition similar to the ``standard" one, \emph{i.e.},  $U_t^{\dagger}=U_t^{-1}$.

For pseudo-hermitian systems an average value of a non-hermitian observable $A$, $\tr{A}$, can be computed as
\begin{equation}
\label{tr}
	\tr{A}  = \sum_{k,\gamma}\bra{\psi_{k,\gamma}} g A\ket{\psi_{k,\gamma}}.
\end{equation}
Formally, this suggests one to use the following Dirac correspondence between bra and ket vectors $\ket{\psi} \leftrightarrow \bra{\psi}g$~\cite{brody2}.

\vspace{0.5cm}
{\noindent \bf Pseudo--hermitian Jarzynski equality}
\vspace{0.5cm}

\noindent
Having analyzed the mathematical structure of pseudo-hermitian quantum systems, we turn to the physical description to analyze the Jarzynski
equality. Without loss of generality and to simplify our notation we assume the spectrum to be non-degenerate.

For an isolated quantum system, the work done during a thermodynamic process $\lambda_t$ of duration $\tau$ is commonly determined by a two-time 
energy measurement~\cite{popescu}. At $t=0$ a projective energy measurement is performed. Next, the system evolves unitarily under the generalized
time-dependent Schr\"odinger equation~(\ref{se}) only to be measured again at $t=\tau$. By  averaging over an ensemble of realizations of such 
processes one can reconstruct the distribution of work values~\cite{tank,pnm},

\begin{equation}
\label{dist}
\mathcal{P}(w) = \sum_{n,m}\delta(w-w_{nm})p_{nm}.
\end{equation}
Above, $p_{nm}$ denotes a probability that a specific transition $\ket{\psi_{n}(\lambda_0)} \rightarrow \ket{\psi_{m}(\lambda_{\tau})}$ will occur, 
whereas $w_{nm}=E_m^{\tau}-E_n$ is the corresponding work done during this transition. It is important to stress that this work is associated with 
$H_t$ rather than $H_t+G_t$ as $G_t$ is a gauge field, and hence it can have no influence on physical observables~\cite{gauge}. 

The transition probability $p_{nm}$ can be seen as the joint probability that the first measurement will yield the energy value $E_n$ given the 
system has been initially prepared in a state $\rho_0$, and the probability that the outcome of the second measurement will be $E_m^{\tau}$ given
the initial state $\psi_n$. Therefore,  

\begin{equation}
\label{p}
p_{nm} = \tr{\Pi_n\rho_0} \times \left |\langle\psi_m^{\tau}, g_{\tau}U_{\tau}\psi_n \rangle\right |^2,
\end{equation}
where $U_{\tau}$  denotes the evolution operator generated by $H_t+G_t$ at time $t=\tau$, whereas $\Pi_n = \langle \psi_n, g_0\,\cdot\, \rangle \psi_n$
is the projector into the space spanned by the $n$th eigenstate. Since $\Pi_n$ is not hermitian the formula for probabilities $p_{nm}$ accounts for the 
metric $g$, and hence   differs from the one usually adopted for hermitian systems~\cite{pnm}.

Assume the system is initially in a Gibbs state , that is $\rho_0=\exp(-\beta H_0)/Z_0$ with $Z_0=\tr{\exp(-\beta H_0)}$ being the partition 
function, then

\begin{equation}
\label{p2}
	 p_{nm}  = \frac{e^{-\beta E_n}}{Z_0}    \langle U_{\tau}^{\dagger} g_{\tau} \psi_m^{\tau}, \Pi_n U_{\tau}^{-1}\psi_m^{\tau} \rangle.
\end{equation}
To obtain the last expression for $p_{nm}$ we have also invoked the unitarity condition~(\ref{gU}). Now, the average exponentiated work can be expressed as

\begin{equation}
\label{dist2}
		 \langle e^{-\beta W} \rangle  = \int dw\mathcal{P}(w)\exp(-\beta w)    
  =\frac{1}{Z_0}\sum_{m,n} e^{-\beta E_m^{\tau}} \langle g_{\tau} \psi_m^{\tau}, U_{\tau} \Pi_n U_{\tau}^{-1}\psi_m^{\tau} \rangle.	 		
\end{equation}
Finally, summing out all projectors $\Pi_n$ and taking into account that $\langle g_{\tau}\psi_m^{\tau},\psi_m^{\tau}\rangle=1$ we arrive at

\begin{equation}
\label{jarz}
 \langle e^{-\beta W} \rangle = \frac{1}{Z_0}\sum_{m}  e^{-\beta E_m^{\tau}} = \frac{Z_{\tau}}{Z_0} = e^{-\beta \Delta F},
\end{equation}
where $F=(-1/\beta)\ln(Z)$ is the system's free energy.

The last equation shows that the Jarzynski equality holds also for non-hermitian systems that admit real spectrum. This is our first main result. 
Jarzynski has shown that the second law of thermodynamics for isothermal processes can be expressed as an equality arbitrarily far from
equilibrium. Our analysis has shown that his result is true for \emph{all non-hermitian systems with real spectrum.}

\vspace{0.5cm}
{\noindent \bf Carnot bound}
\vspace{0.5cm}

\noindent
In the preceding section we argued that if the two-time energy measurement can be performed on a non-hermitian quantum system, then
the Jarzynski equality holds as long as the eigenenergies are real. Now, we will prove that the Carnot statement of the second law 
is also true for all pseudo-hermitian systems.

Consider a generic system that operates between two heat reservoirs with hot, $T_h$,  and cold, $T_c$, temperatures, respectively. 
Then, the Carnot engine consists of two isothermal processes during which the system absorbs \emph{or} exhausts heat and two thermodynamically
adiabatic, that is, isentropic strokes while the extensive control parameter $\lambda$ is varied~\cite{carnot,Gong15}. It is well established
that the maximum efficiency $\eta$ for classical systems, attained in the quasistatic limit, is given by the Carnot 
bound~\cite{carnot_24,Huang12,Long15}:

\begin{equation}
\label{eta}
\eta = 1 - \frac{T_c}{T_h} < 1.
\end{equation}
Recent years have witnessed an abundance of research~\cite{Scovil_1967,Lutz_2014,Scully_2003} investigating whether quantum correlations
can be harnessed to break this limit. Recently, the Carnot limit has been proven to be universal within the usual framework~\cite{carnot}. 
This limit can be seen as yet another formulation of the second law of thermodynamics for quasistatic processes. We will show that is holds
for all pseudo-hermitian systems whether their spectrum is real or not.

We begin by proving that both the energy $E=\tr{\rho H}$ and entropy $S$ are real in our present framework. Indeed, from~(\ref{eq04}) it 
immediately follows that

\begin{equation}
\label{real}
E^*=\tr{g\rho g^{-1} gH g^{-1}} = E,
\end{equation}
with $\rho$ being a Gibbs thermal state. Interestingly, this result holds true even if some of the eigenvalues $E_n$ are
complex. Note, in that case $g$ exists but is not positive definite and thus cannot be expressed like in Eq.~(\ref{eq05}).

To understand why Eq.~(\ref{real}) holds when complex eigenvalues appear in conjugate pairs note that
$\ket{\psi_{n,\alpha}}=g^{-1}\ket{\phi_{n,\alpha}}$, and consider

\begin{equation}
H\ket{\psi_{n,\alpha}}= g^{-1}H^{\dagger}\ket{\phi_{n,\alpha}} = E_n^*\ket{\psi_{n,\alpha}},
\end{equation}
showing that if $E_n$ is in the spectrum of $H$ so is $E_n^*$. Moreover $g^{-1}$ maps the subspace spanned by all eigenvectors 
belonging to $E_n$ to that belonging to $E_n^*$. Since $g^{-1}$ is invertible, the mapping is one-to-one, and the multiplicity
of both $E_n$ and $E_n^*$ is the same. An interesting realization of such systems is the non-hermitian tight-binding model~\cite{Hatano96}.
 
The result~(\ref{real}) can also be obtained directly, that is, without invoking the metric $g$ explicitly. Indeed, we have

\begin{equation}
\label{direct}
E = \tr{\rho H} = \frac{1}{Z}\sum_{n} E_n e^{-\beta E_n} 
                = -\frac{1}{2Z}\frac{\partial}{\partial\beta}\sum_{n/2} \left( e^{-\beta E_n} + e^{-\beta E_n^*}\right)=E^*.
\end{equation}

In the present case, the thermodynamic entropy is given by the von Neumann entropy~\cite{Sagawa_2015}. The latter can be further simplified
and it takes the well known form $S=\beta (E - F)$~\cite{carnot}. Since the partition function $Z$ is real so is the free energy $F$. Hence,
we conclude that the entropy $S$ is real.

According to the first law of thermodynamics~\cite{callen}, $dE=\delta Q + \delta W$,
%
%
there are two forms of energy: heat $\delta Q$ is the change of internal energy associated with a change of entropy, whereas work 
$\delta W$ is the change of internal energy due to the change of an extensive parameter, \emph{i.e.}, change of the Hamiltonian of
the system. To identify those contributions we write~\cite{carnot}
\begin{equation}
\label{1law}
 	   		\td E  = \tr{\delta\rho\, H} + \tr{\rho\, \delta H}.           
\end{equation}
In the quasistatic regime, the second law of thermodynamics for isothermal processes states that $dS=\beta \delta Q$. Combining the latter 
with~(\ref{1law}) proves that (i) $\delta Q$ and thus $\delta W$ are real and (ii) the intuitive definitions of heat and work introduced in~\cite{alicki_79} apply also to pseudo-hermitian systems. 

After completing a cycle, a quantum pseudo-hermitian heat engine has performed work $\langle W \rangle =\langle Q_h\rangle-\langle Q_c \rangle$
and exhausted a portion of heat $\langle Q_c \rangle$ to the cold reservoir. Therefore, the efficiency of such a device is given by~\cite{carnot}

\begin{equation}
\label{eta2}
\eta = \frac{\langle W \rangle}{\langle Q_c \rangle} = 1 - \frac{T_c}{T_h}.
\end{equation}
In conclusion, we have shown that the Carnot bound, which expresses the second law of thermodynamics for quasistatic processes, holds for all 
pseudo-hermitian systems. In contrast,  the second law for arbitrarily fast processes encoded in the Jarzynski equality~(\ref{jarz}), \emph{only}
holds for all non-hermitian systems with real spectrum.

\vspace{0.5cm}
{\noindent \bf \Large Discussion}
\vspace{0.5cm}
\noindent

{\noindent \bf Example 1a}
\vspace{0.5cm}
\noindent

We begin with a model for localization effects in solid state physics~\cite{Hatano96}. 
The general form of its Hamiltonian in one dimension reads

\begin{equation}
\label{loc}
H = \frac{(p-i\xi)^2}{2m} + V(x),
\end{equation} 
where $V(x)$ is a confining potential, and $p$ and $x$ are the momentum and position operators respectively. 
They obey the canonical commutation relation $\left[x,p\right]=i\hbar$. Real parameter $\xi$ expresses an 
external magnetic field and $m$ is the mass. Using the Baker-Campbell-Hausdorff formula one can verify that 

\begin{equation}
\label{BH}
e^{2\xi x}  p e^{-2\xi x} = p + 2\xi \left[x,p\right] + 2\xi^2 \left[x, \left[x,p\right]\right] + \dots = p + 2i\xi.
\end{equation}
Therefore, since $[V(x),e^{2\xi x}]=0$, we conclude that $H$ is pseudo-hermitian. The metric $g=e^{2\xi x}$ is positive
definite and thus the spectrum of~(\ref{loc}) is real. 
Further, we assume that the corresponding classical potential $V_{\text{c}}(x)$ has a non-vanishing second derivative, 
and a minimum at $x=0$ (\emph{e.g.} $V_{\text{c}}'(0)=0$). Then

\begin{equation}
\label{pot}
V_{\text{c}}(x) = V_{\text{c}}'(0)x + \frac{1}{2}V_{\text{c}}''(0)x^2 + O(\delta x^3) \approx  \frac{1}{2}m\omega^2 x^2,
\end{equation}
where $V_{\text{c}}'(0)=\frac{1}{2}m\omega^2$ has been introduced. After quantization, the eigenvalues and eigenvectors 
of this non-hermitian harmonic oscillator read (for the sake of simplicity we set $m=\hbar=1$ throughout)

\begin{equation}
\label{ic}
 \psi_n(x): = \frac{1}{\sqrt{2^n n!\sqrt{\pi}}} H_n(\sqrt{\omega}x) e^{-\omega x^2-\xi x},
 \quad
 E_n = \omega\left(n+\frac{1}{2}\right),
\end{equation}
where $H_n(x)$ are the Hermite polynomials.

Now we assume that the size of this harmonic trap (\emph{e.g.} $\omega$) is changed, and thus $g$ does not depend on time.
Experimentally, harmonic traps are sensitive to initial excitations resulting for a discontinuity of the protocol itself
at the beginning~\cite{Singer13}. The most common way to minimize this effect, while quenching between $\omega_i$, and 
$\omega_f$, is to use functions smooth enough at the ``edges'', for instance,

\begin{equation}
\label{erf}
\omega(t) = \frac{\omega_i+\omega_f}{2} + \frac{\omega_i-\omega_f}{2} \text{erf}(t/\tau),
\quad
-N\tau < t < N\tau
\end{equation}
where $\text{erf}(\cdot)$ denotes the error function, $\tau$ is a time scale, and $N$ is an integer emulating
infinity. The transition probabilities~(\ref{p}) can be expressed via the following integral

\begin{equation}
\label{hp}
p_{nm} = \frac{\exp\left(-\beta\omega_i(n+1/2)\right)}{\sinh\left(\beta\omega_i/2\right)} 
\int_{-\infty}^{\infty} e^{2\xi x}\, \psi_n^{N\tau}(x)^*\,\psi_m(x,N\tau) \, dx,
\end{equation} 
where the partition function $Z_0=1/\sinh\left(\beta\omega_i/2\right)$ has been calculated exactly; and
$\psi_m(x,N\tau)=U_{N\tau}\psi_n(x)$ is the solution of Eq.~(\ref{se}), with the initial condition given
by~(\ref{ic}), at $t=N\tau$. Although $\psi_m(x,N\tau)$ cannot be obtained analytically, a closed form 
expressed in terms of a solution to the corresponding classical equation of motion can be found (see 
\emph{e.g.}~\cite{Youhong98}).
\begin{figure}
	\includegraphics[width=\textwidth]{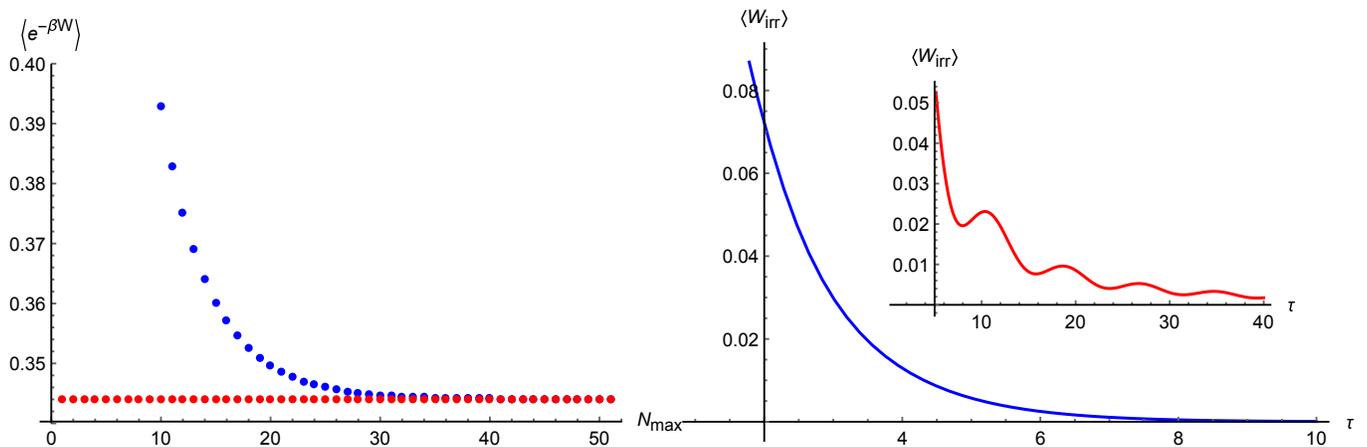}
	\caption{\label{fig:ho}
		Left panel: 
		Average exponentiated work $\langle e^{-\beta W}\rangle$ (blue curve) as a function of the number of terms $N_{\text{max}}$
		included in the summation~(\ref{dist2}) for the protocol~(\ref{erf}). The function quickly converges to $e^{-\beta\Delta F}$ 
		(red curve) showing that the Jarzynski equality~\eqref{jarz} holds. 
		Right panel: 
		$\langle W_{\text{irr}}\rangle=\langle W \rangle - \Delta F$ as a function of $\tau$ which relates to the speed at which the 
		energy is supplied to the system. The irreversible work $\langle W_{\text{irr}}\rangle\to 0 $ as $\tau$ approaches the quasistatic
		regime.  The inset (red curve) shows the irreversible work calculated for a linear protocol, $\omega(t) = \omega_i + (\omega_f - \omega_i) t/\tau$. We see that it takes longer for the system to reach its quasistatic regime. Parameters used in the numerical simulations are: $w_i=0.2$, $w_f=0.6$, $N\tau=1.5$ (left panel) and $N\tau=3.$ (right panel); the remaining parameters were set to $1$.
	}
\end{figure}

Figure~\ref{fig:ho} (Left panel) shows the average exponentiated work $\langle e^{-\beta W}\rangle$ (blue curve) as a function of the 
number of terms $N_{\text{max}}$ included in the summation~(\ref{dist2}). This function quickly converges to $e^{-\beta\Delta F}$ 
proving that the Jarzynski equality~\eqref{jarz} holds. 
On the right panel we have depicted the irreversible work $\langle W_{\text{irr}}\rangle=\langle W \rangle - \Delta F$ (blue curve) as a function
of $\tau$ which determines the speed at which the energy is supplied to the system. When $\tau\to\infty$ the system enters its quasistatic
regime and the irreversible work becomes negligible, that is $\langle W_{\text{irr}}\rangle\to 0$~\cite{Deffner08, Lutz09}. The inset (red 
curve) shows the irreversible work calculated for a linear protocol, $\omega(t) = \omega_i + (\omega_f - \omega_i) t/\tau$. As we can see, 
it takes longer for the system to reach its quasistatic regime. Moreover, the oscillatory behavior is a signature of the initial excitation
which dominates for fast quenches (small $\tau$).

\vspace{0.5cm}
{\noindent \bf Example 1b}
\vspace{0.5cm}
\noindent

Another class of systems that is used to explain localization effects relates to non-hermitian tight-binding models~\cite{Longhi10,Longhi13}. 
For example
\begin{equation}
H = -\frac{t}{2}
\sum_{\mathbf{x}} \sum_{\nu=1}^{d} e^{\mathbf{\alpha} \cdot \mathbf{e_{\nu}}} a_{ \mathbf{x} + \mathbf{e_{\nu}}}^{\dagger} a_{\mathbf{x}}
+  e^{\mathbf{-\alpha} \cdot \mathbf{e_{\nu}}} a_{\mathbf{x} }^{\dagger} a_{\mathbf{x} + \mathbf{e_{\nu}}}
+\sum_{\mathbf{x}} V_{\mathbf{x}} a_{ \mathbf{x}}^{\dagger} a_{\mathbf{x}},
\end{equation}
where, $a_{ \mathbf{x}}^{\dagger} $ and $a_{\mathbf{x}}$ are bosonic creation and annihilation operators respectively,
$\mathbf{e_{\nu}}$ are the unit lattice vectors, and $t$ is the hopping parameter, and $ V_{\mathbf{x}}$ denotes the
on-site potential. Interestingly, the complex eigenvectors appear in conjugate pairs (see Eq.~($2$) in~\cite{Hatano96}
and the discussion that follows). Therefore, this model provides another example for a building block of a non-hermitian 
Carnot engine.

\vspace{0.5cm}
{\noindent \bf Example 2}
\vspace{0.5cm}
\noindent

The remainder of the present work is dedicated to a careful study of a second, experimentally relevant example~\cite{gao}.
Consider a two level system described by the Hamiltonian

\begin{equation}
\label{ex1}
H_t =
%
\lambda_t \sigma_{+}\sigma_{-} + \lambda_t^* \sigma_{-}\sigma_{+} +  \gamma \sigma_{+} +  \gamma^* \sigma_{-},
\end{equation}
where $\lambda_t$ is a complex control parameter, and $\gamma$ is a complex constant, whereas $\sigma_{+}$ and $\sigma_{-}$ are the
raising and lowering fermionic operators. This simple model~(\ref{ex1}) has been extensively studied in the literature~\cite{Deffner2015a,bender_1, bender_2}, and it has been also realized experimentally both in optics~\cite{optic} and semiconductor microcavities~\cite{gao}. 

\begin{figure}
	\includegraphics[width=\textwidth]{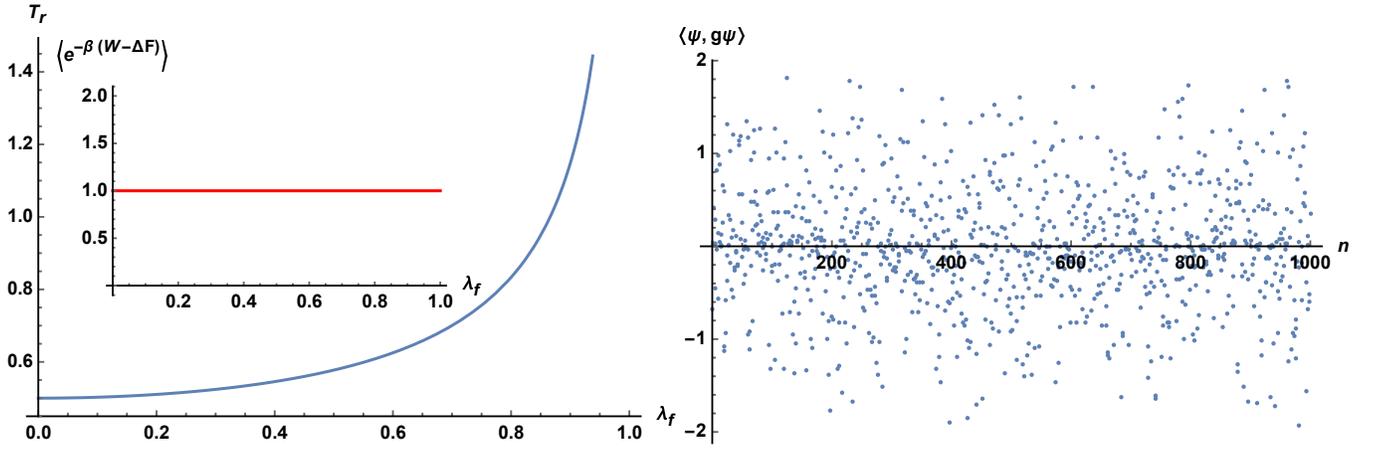}
	\caption{\label{fig:jarz} Left panel: Relaxation time $T_r=|E_1^{\tau}-E_2^{\tau}|^{-1}$, as a function of the final value $\lambda_f$ for the 
		linear quench $\lambda_t = \lambda_i + (\lambda_f - \lambda_i) t/\tau$. Parameters are $\lambda_i=0$, $\beta=\hbar=\tau=1$. Inset: numerical confirmation of the Jarzynski equality~\eqref{jarz}. Right panel: In the broken regime quantum work can no longer be determined by the two-time
		energy measurement as $\langle\psi,g\psi\rangle$ can be both positive and negative. To construct the plot we set $g=\sigma_x$. 
		States $\psi(n)$ have been chosen randomly; and $n$ is an integer that has been assigned to them.
	}
\end{figure}

To make the spectrum of~(\ref{ex1}) real  we set $\lambda_t$ to be purely imaginary ($\lambda_t\rightarrow i\lambda_t$); and without any loss of generality
we choose $\gamma=1$. This corresponds to the following parameters $E_{1,2}=0$, $\Gamma_{1,2}=\pm\lambda_t$, and $q=\gamma=1$ for the hybrid light--matter system of quasiparticles investigated in~\cite{gao}. Such systems are formed as a result of a strong interaction between excitons and photons in a semiconductor microcavity~\cite{ep2}. They are commonly referred to as exciton--polaritons~\cite{ep}.

A simple calculation shows that $H_t^{\dagger}=\sigma_xH_t\sigma_x$, where $\sigma_x$ is the Pauli matrix in $x$ direction. Thus $H_t$ is
indeed pseudo-hermitian. However, the corresponding $\sigma_x$ is not a metric. For instance $\langle e_1,\sigma_xe_1\rangle=0$, where 
$\ket{e_1}=(1,0)^t$.  Nevertheless, we can easily find one by rewriting $H_t$ in its diagonal form,

\begin{equation}
V_t^{-1}H_tV_t  =
\begin{pmatrix}
E_1^t & 0 \\
0 & E_2^t 
\end{pmatrix},
\quad
E_{1,2}^t = \pm \sqrt{1-\lambda_t^2}.
\end{equation}
Note, both $E_{1,2}^t$ are real as long as $\lambda_t \le 1$, otherwise $E_1^t=E_2^{t*}$. Therefore, the Carnot 
bound~(\ref{eta}) holds in both these regimes, whereas the Jarzynski equality~(\ref{jarz}) only in the first one.
Now, the proper metric can be defined via the similarity transformation $V_t$

\begin{equation}
\label{gt}
g_t = V_t^{\dagger}V_t =  
2
\begin{pmatrix}
1 & - i \lambda_t \\
i \lambda_t &1 
\end{pmatrix}.
\end{equation}  

To investigate the dynamics of~(\ref{ex1}) we assume that $\lambda_t$ changes on a time scale $\tau$ in a linear manner, that is
$\lambda_t = \lambda_i + (\lambda_f - \lambda_i) t/\tau$. The linearity does not pose any restriction on our analysis as the 
Jarzynski equality holds for all protocols $\lambda_t$~\citep{Deffner2015a}. Figure~\ref{fig:jarz} (Left panel) depicts the 
relaxation time $T_r=\Delta^{-1}$, where $\Delta=|E_1^{\tau}-E_2^{\tau}|$, as a function of the final value $\lambda_f$~\cite{Gogolin_2015}. 
The relaxation time diverges as $\lambda_f$ approaches the critical point at $\lambda=1$. Similar behavior has been observed for the 
irreversible work $\langle W_{\text{irr}} \rangle := \langle W \rangle - \Delta F$ in $\mathcal{P}\mathcal{T}$-symmetric systems~\cite{Deffner2015a}. 
The critical point separates the unbroken domain, where energies are real, from the broken one characterized by complex energy values. 
The energetic cost associated with a potential crossover between those two regimes becomes infinite, and the system ``freezes out" before
even having a chance to cross to the other regime~\cite{kz2,kz}. 

In the broken regime, Eq.~(\ref{gt}) no longer reflects pseudo-hermiticity of the system, that is $V_t$ does not fulfill Eq.~(\ref{eq04}).
In fact, all operators $g$ for which the latter equation is true, $\sigma_x$ being an example (see Fig.~\ref{fig:jarz}, Right panel), lead to 
indefinite inner product spaces. Note that in Fig.~\ref{fig:jarz} (Right panel) the norm can be both positive and negative. Therefore, the
evolution within those spaces cannot be unitary and the two-time energy measurement paradigm can no longer be applied~\cite{paolo}. In the
quasistatic limit, however, quantum work can still be defined, and we have shown that the second law still holds for all pseudo-hermitian systems. 

\vspace{0.5cm}
{\noindent \bf Conclusions}
\vspace{0.5cm}
\noindent

In summary, we have carefully studied thermodynamic properties of quantum systems that do not satisfy one of the basic requirements imposed on them by 
the axiom of quantum mechanics - hermiticity. We have shown that if quantum work can be determined by the two-time projective energy measurements, 
then the Jarzynski equality still holds for non-hermitian systems with real spectrum. Note, this equality expresses the second law of thermodynamics 
for isothermal processes arbitrarily far from equilibrium.

We have also argued that the Carnot bound is attained for all pseudo-hermitian systems in the quasistatic limit. Furthermore, we have 
also proposed an experimental setup to test our predictions. As elaborated in the previous section, the system in question consists of 
strongly interacting excitons and photons in a semiconductor microcavity~\cite{gao}. Moreover, we have investigated two non-hermitian
models that where originally introduced to explain localization effects in solid state physics~\cite{Hatano96}. First one, a non-hermitian
harmonic oscillator that admits real spectrum was used to demonstrate the Jarzynski equality. The second one, the so called non-hermitian 
tight-binding model was given as an example of a quantum system having complex eigenenergies that appear in conjugate pairs. This model
provides another example of a building block of a non-hermitian Carnot engine.

\vspace{0.5cm}
{\noindent \bf \Large References}
\vspace{0.5cm}

%
%

\vspace{0.5cm}
{\noindent \bf \Large Acknowledgments}

\vspace{0.5cm}
\noindent
This work was supported by the Polish Ministry of Science and Higher Education under project Mobility Plus 1060/MOB/2013/0 (B.G.); 
S.D. acknowledges financial support from the U.S. Department of Energy through a LANL Director's Funded Fellowship. 


\vspace{0.5cm}
{\noindent \bf \Large Author contributions}

\vspace{0.5cm}
\noindent
B.G., S.D. and A.S developed ideas and derived the main results. B.G prepared figures $1$ and $2$. B.G., S.D. and A.S. wrote
and reviewed the manuscript.

\vspace{0.5cm}
{\noindent \bf \Large Competing financial interests}

\vspace{0.5cm}
\noindent
The authors declare no competing financial interests.

\end{document}